# Thermodynamic Relaxation Drives Expulsion in Giant Unilamellar Vesicles

C.T. Leirer[1,2], B. Wunderlich[1], A. Wixforth[1], M.F. Schneider[1*]

[1] University of Augsburg, Experimental Physics I, D-86159 Augsburg, Germany
[2] Niels Bohr Institute, University of Copenhagen, DK - 2100 Copenhagen Ø, Denmark

*Corresponding author:*

Matthias F. Schneider

Phone: +49-821-5983311, Fax: +49-821-5983227

<matthias.schneider@physik.uni-augsburg.de>

University of Augsburg, Experimental Physics I,

Biological Physics Group

Universitätstr. 1

D-86159 Augsburg, Germany

**Abstract**

We investigated the thermodynamic relaxation of giant unilamellar vesicles (GUVs) which contained small vesicles within their interior. Quenching these vesicles from their fluid phase ($T>T_m$) through the phase transition in the gel state ($T<T_m$) drives the inner vesicles to be expelled from the larger mother vesicle via the accompanying decrease in vesicle area by ~25% which forces a pore to open in the mother vesicle. We demonstrate that the proceeding time evolution of the resulting efflux follows the relaxation of the membrane area and describe the entire relaxation process using an Onsager-like nonequilibrium thermodynamics ansatz. As a consequence of the volume efflux internal vesicles are expelled from the mother vesicle. Although complete sealing of the pore may occur during the expulsion, the global relaxation dynamics is conserved. Finally, comparison of these results to morphological relaxation phenomena found in earlier studies reveals a universal relaxation behaviour in GUVs. When quenched from the fluid to gel phase the typical time scale of relaxation shows little variation and ranges between 4-5 seconds.

**Introduction**

The traffic and transport of small vesicles or macromolecules inside or across the membrane of the cell, organelles or liposomes has gained new attention since evidence has been given that elasticity is an important player in understanding cell or vesicle adhesion and trafficking [1] [2] [3]. For instance, it has been shown, that the mechanical properties of biological membranes can be modified locally by the adsorption of protein coats which in turn induce endocytosis or exocytosis [3] [4]. Using lipid mixtures which exhibit a phase transition in an easily controllable temperature range, allows the manipulation of both the area/volume ratio and the vesicle compressibility. Both area and compressibility increase ("softening") during lipid chain melting.

In this paper, we demonstrate how the lipid phase transition serves to control the expulsion of internalized vesicles from the interior of the mother vesicle ("birthing of vesicles"). We show that the expulsion process is a consequence of the thermodynamic relaxation after quenching from the fluid to the gel phase and can be readily described in an Onsager-like treatment. It appears that the relaxation follows a universal behaviour throughout all our experiments including morphological transitions reported elsewhere [Leirer et al].

*Materials and Methods*

1,2-Dipalmitoyl-*sn*-Glycero-3-Phosphocholine (DPPC) dissolved in chloroform (20mg/ml) was purchased from Avanti Polar Lipids (Alabaster, Alabama, USA) and was used without further purification. Vesicle electroformation was performed as described elsewhere [5, 6]. We prepared all aqueous solutions with ultrapure water (pure Aqua, Germany) with a specific resistance of 18.2MΩ. Vesicles were prepared by spreading a small amount of lipid in chloroform on an ITO coated glass substrate and left it in a vacuum to remove any traces of the organic solvent. Subsequently, an AC field of 1V/mm and 10Hz was applied between the conducting ITO slides, forming the swelling chamber. The assembled chamber was fixed and placed in a thermal water bath at 50°C. After 3 hours the electroswelling process was completed and Giant Unilamellar Vesicles with diameters of up to 100μm were harvested for use. Fluorescence microscopy of the vesicles was enabled by adding 0.1% of the fluorescent dye Texas Red (Invitrogen, Karlsruhe, Germany). Experiments were performed in a temperature controllable chamber with optical access from the bottom and top. Fluorescent images were collected with a standard CCD camera (Hamamatsu Photonics Deutschland, Herrsching am Ammersee, Germany) coupled to an Axiovert 200 microscope (Zeiss, Oberkochen, Germany) and the temperature was controlled with the aid of a standard heat bath (Julabo, Seelbach, Germany).

**Results**

*Phase transition induced pores*

The typical expulsion scenario observed when quenching a GUV with inner vesicles from the fluid phase through the phase transition region into the gel phase, is shown in Fig. 1. Upon cooling, the inner vesicle moves towards the wall of the mother vesicle, as a consequence of volume efflux and hydrodynamic coupling and starts to exit the GUV through a pore. The pore is generated at the beginning of the process due to the emerging tension and the incompressibility of water. The external bath cooling the vesicles is basically macroscopic, therefore supplying an infinite amount of energy to complete the phase transition of the lipid vesicle. Due to the accompanied total decrease in area by ~ 25% the membrane is forced to shrink. Furthermore, due to the fast cooling rate of 10°C/s, the lipid bilayer appears impermeable to water and the process can be considered to take place under constant volume constraints. However, since the vesicle is spherical and rather taut prior to cooling, it has already reached its maximum volume/area ratio. Consequently, the only way to shrink the surface of the mother vesicle is the expulsion of volume after creating a pore at a tension $\sigma_{lysis} \approx 1 mN/m$ [7]. The actual nucleation centre of the pore is most likely a membrane defect. In contrast to Bar-Ziv et al [8] [9] and Menger et al [10], who observed similar expulsion phenomena with the driving force being of chemical or osmotic origin, here the lysis tension is created by the forced area decrease during phase transition at constant volume constrain. For a fluid-like vesicle the surface tension drops after opening of the pore for two reasons [11]: i) the membrane area relaxes and becomes wrinkled and ii) the leak-out of liquid due to the excess Laplace pressure, which in turn is a consequence of the residual tension left in the vesicle membrane due to the pore line tension. Here, however, the vesicle is in its gel-like phase where lateral shear viscosity impairs rapid pore closure. After the vesicle is expelled (lower row in Fig. 1) the pore does not reseal but remains at a rather constant diameter for several seconds after expulsion (see inset in Fig. 1). The area of the pore $A_p$ observed after expulsion is typically in the range of 1-2% of the entire area of the gel-like vesicle $A_g$. To calculate the critical area extension $\Delta A_{crit}$ leading to membrane rupture we need to compare the elastic energy of the membrane with

$$F_P = \frac{1}{2}\kappa_A^T A \left( \frac{\Delta A}{A} - \frac{\pi r^2}{A} \right)^2 + 2\pi r \gamma \qquad (1)$$

and without a pore [12]

$$F_V = \frac{1}{2}\kappa_A^T \frac{(\Delta A)^2}{A} \qquad (2)$$

Where $\kappa_A^T$ represents the isothermal lateral compressibility, $\gamma$ the line tension and $r$ the radius of the pore. The moment the pore opens, the tension in the membrane and the tension created by the pore are the same, i.e.

$$\frac{1}{2}\kappa_A^T \frac{(\Delta A_{crit})^2}{A} \approx 2\pi r\gamma \qquad (3)$$

In this case, the change in area to relax the membrane is exactly the area of the pore and hence

$$\frac{1}{2}\kappa_A^T \frac{(\Delta A_{crit})^2}{A} \approx 2\pi\gamma\sqrt{\frac{\Delta A_{crit}}{\pi}} \qquad (4)$$

, which finally leads to

$$\Delta A_{crit} = 4A^{2/3}\pi^{2/5}\left(\frac{\gamma}{\kappa_A^T}\right)^{2/3} \qquad (5)$$

For typical numbers $\kappa_A^T \approx 10^{-1} N/m$ and $\gamma \approx 10^{-12} N$ this leads to $\Delta A_{crit} \approx 0{,}01\ A_g$. As the critical area extension $\Delta A_{crit}$ is almost completely released upon opening of a pore, the area extension can be compared to the initial pore size and is in very good agreement with the experimental observations.

*Relaxation and Expulsion*

After pore opening the membrane continues to shrink until the entire membrane is in the gel-state. In the following we demonstrate that the volume efflux depends only on the change in area. We describe the relaxation process driving the volume to be expelled using an Onsager-type relation [13], which according to Grabitz et al. [14] results in a single exponential decay of the enthalpy difference

$$H - \overline{H}(T) = H - \overline{H}(0)\exp\left(-\frac{t}{\tau}\right)$$

where $H - \overline{H}(0)$ is the initial difference in enthalpy and $\tau$ the typical time scale of the relaxation. Employing the proportionality of the excess enthalpy and area [15]

$$\Delta A(T) = \alpha \Delta H(T)$$

with $\alpha$ being constant, within experimental errors, for phosphocholines and lipid mixtures, we can express the flux of area after a disturbance back to equilibrium

$$\Delta A(T) = \Delta A_{in} \cdot \exp\left(-\frac{t}{\tau}\right)$$

where $\Delta A_{in}=A_f - A_g=0.25 \cdot A_g$ is the difference between the area in the fluid ($A_f$) and the gel phase ($A_g$). Consequently the total area obtained is

$$A(T) \approx A_g\left[1+\frac{1}{4}\exp\left(-\frac{t}{\tau}\right)\right] \qquad (3)$$

The actual efflux of water $Q$, is related to the change in volume by

$$Q \approx -\frac{dV}{dt}. \qquad (4)$$

For a spherical vesicle the volume is related to the actual area by $V(t) = 1/6/\sqrt{\pi}\,A(t)$ which leads to a time course for the volume efflux given by

$$Q \approx \frac{1}{16\sqrt{\pi\tau}}\left(A_g\right)^{3/2} \cdot \exp\left(-\frac{t}{\tau}\right)\left[1+1/4\exp\left(-\frac{t}{\tau}\right)\right] \qquad (5)$$

In Fig. 2a the evolution of the area after the temperature quench is displayed. Subsequently to the pore opening at $t=3s$ (Fig. 2a) the global area relaxation is well described by Eq. 3. We obtain a typical relaxation time of $5.0 \pm 0.3$ s. The precise description of the averaged (smoothed) volume efflux by Eq. 5 (inset Fig. 2a) demonstrates that it is exclusively driven by the area relaxation. The typical relaxation times found throughout our experiments, ranged between $\tau \approx 4 - 5$ s and are in the same order as morphological transitions described elsewhere [16]. When objects larger then the pore size are expelled almost complete sealing takes place (Fig. 2b). Therefore, when taking the fine steps in area relaxation (Fig. 2a) into account when calculating the volume efflux $Q$, several maxima and minima appear (Fig. 2b) whenever the area is hindered in its relaxation.

Of course, the efflux is driven by a pressure difference across the membrane which results from the Laplace pressure $\Delta P$

$$\Delta P = \frac{2\sigma}{R(t)} \qquad (6)$$

where $\sigma$ is the surface tension and $R(t)$ the time dependent radius of the vesicle. The actual efflux $Q$ is related to the pressure drop by

$$Q = \frac{\Delta P}{3\eta} r^3$$

where $\eta$ is the viscosity of water and stokes friction is assumed. It is obvious that the pore radius $r(t)$ and $R(t)$ has to follow the exact same relaxation process as the membrane (Eq. 3). Therefore the Laplace pressure decays in the same manner as the area relaxes, unifying the thermodynamic and mechanical point of view.

**Conclusion**

In summary, we here reported on thermodynamic controlled expulsion of small internalized vesicles from a mother vesicles by driving the system from the fluid phase ($T>T_m$) through its phase transition into the gel phase ($T<T_m$). The drastic change in area at constant volume drives the formation of pores in the mother vesicle. We demonstrate that the time course of vesicle or volume expulsion is driven by the thermodynamic relaxation of the membrane and can be described by a simple non-equilibrium Onsager-like relation.

Finally, bearing in mind, that the lipid phase transition can also be induced by proteins, pressure or chemically produced pH-gradients, this process may be a powerful mechanism in controlling membrane trafficking.

**Acknowledgement**

Financial support by the Deutsche Forschungsgemeinschaft DFG (SFB 486 and SCHN 1077), the Fond of the Chemical industry, Elite Netzwerk Bayern (CompInt) and the German Excellence Initiative via the Nanosystems Initiative Munich (NIM) is acknowledged. CTL was supported by the Bayerische Forschungsstiftung.

**Figures and Figure Captions**

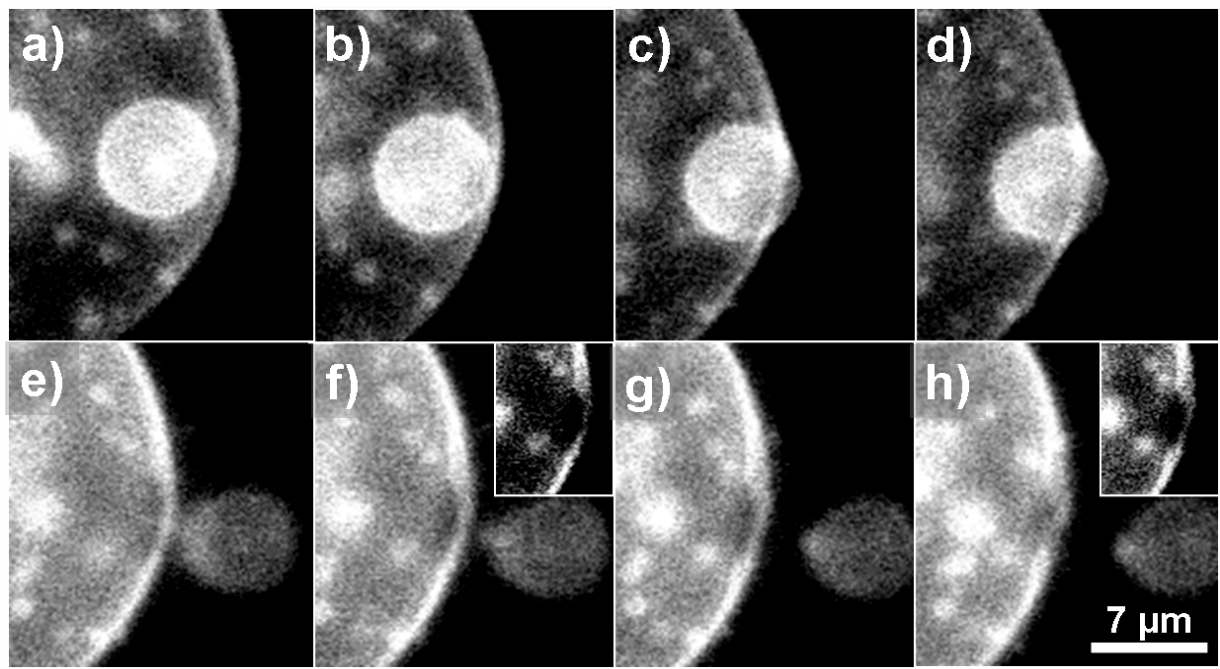

**Figure 1:** DPPC vesicle after temperature quench from the fluid to the gel phase. When a pore opens the enclosed vesicle approaches the wall of the mother vesicle and seals the pore. The raising pressure deforms the membrane (c, d) and eventually expels the vesicle through the pore (e-h). The inset of images f and h show that the pore remains open after vesicle expulsion.

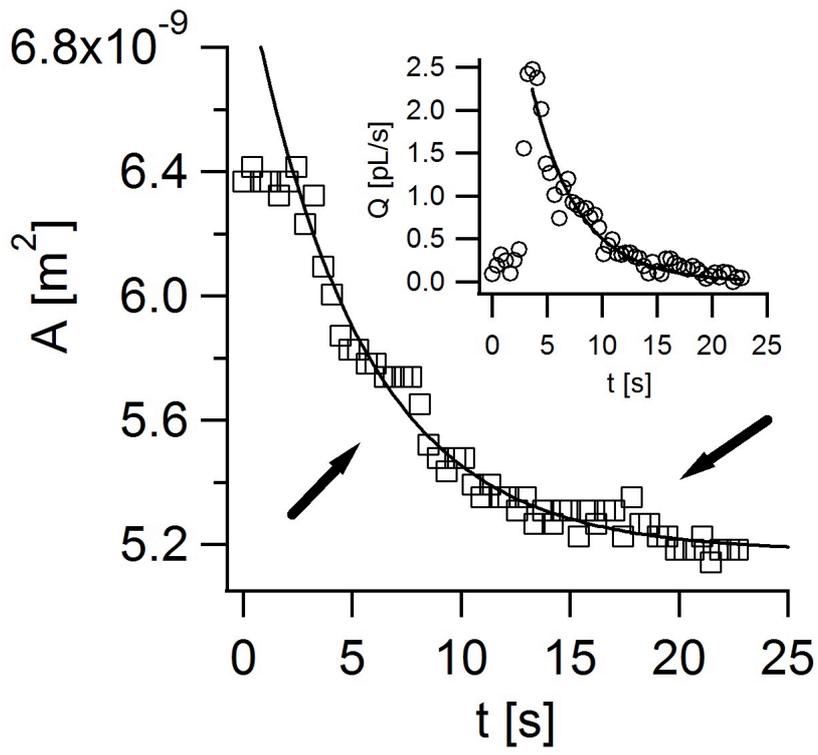

**Figure 2a**

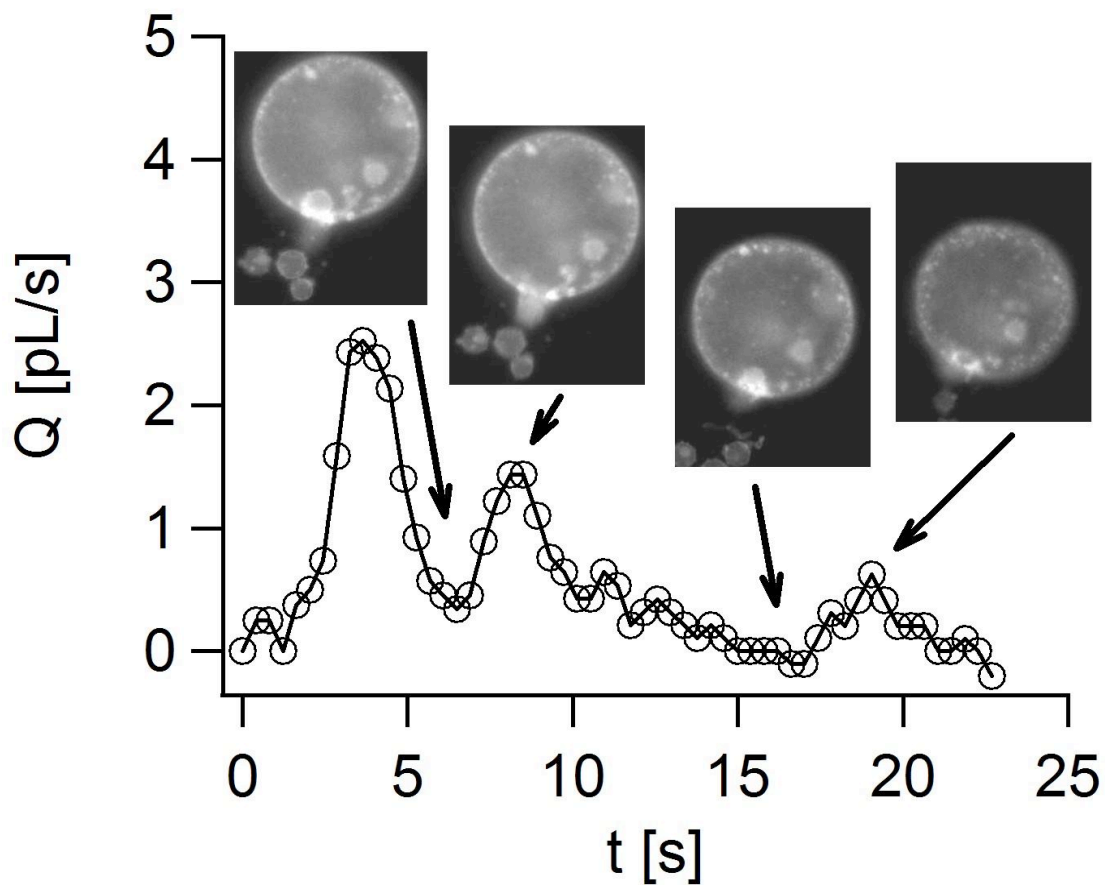

**Figure 2b**

**Figure 2:** a) Time evolution of the area of a DPPC vesicle after temperature quench from the fluid to the gel phase and consequent pore opening. The global exponential decay provides a relaxation time of 5.0 ± 0.3 s. The volume efflux (inset) calculated form the averaged (smoothed) area decay which is readily described by Eq. 5. Steps in the area relaxation (arrows) indicate sealing of the pore due to the expulsed vesicles. b) When calculating $Q$ from the actual area relaxation of figure 2a (not from the smoothed curved) these steps are represented as spikes in the volume efflux.